\documentclass[twocolumn,final,balancelastpage,tightenlines,                             amsmath,aip,superscriptaddress,a4paper]{revtex4}

\usepackage[dvipdfm]{graphicx}
\usepackage{latexsym}
\begin{document}
\title{%
Tuning of metal-insulator transition of 
two-dimensional electrons at parylene/SrTiO$_3$ interface
by electric field%
}
\author{H. Nakamura}
\affiliation{Department of Advanced Materials,
	University of Tokyo, Kashiwa 277-8561, Japan.}
\affiliation{Correlated Electron Research Center (CERC),\\
	National Institute of Advanced Industrial Science and Technology (AIST),\\
	AIST Tsukuba Central 4, Tsukuba 305-8562, Japan}
\author{H. Tomita}
\affiliation{Department of Advanced Materials,
	University of Tokyo, Kashiwa 277-8561, Japan.}
\affiliation{Correlated Electron Research Center (CERC),\\
	National Institute of Advanced Industrial Science and Technology (AIST),\\
	AIST Tsukuba Central 4, Tsukuba 305-8562, Japan}
\author{H. Akimoto}
\affiliation{Advanced Science Institute, RIKEN (The Institute of Physical and
Chemical Research), Wako 351-0198, Japan}
\author{R. Matsumura}
\affiliation{Department of Advanced Materials,
	University of Tokyo, Kashiwa 277-8561, Japan.}
\author{I. H. Inoue}
\affiliation{Correlated Electron Research Center (CERC),\\
	National Institute of Advanced Industrial Science and Technology (AIST),\\
	AIST Tsukuba Central 4, Tsukuba 305-8562, Japan}
\author{T. Hasegawa}
\affiliation{Correlated Electron Research Center (CERC),\\
	National Institute of Advanced Industrial Science and Technology (AIST),\\
	AIST Tsukuba Central 4, Tsukuba 305-8562, Japan}
\author{K. Kono}
\affiliation{Advanced Science Institute, RIKEN (The Institute of Physical and
Chemical Research), Wako 351-0198, Japan}
\author{Y. Tokura}
\affiliation{Correlated Electron Research Center (CERC),\\
	National Institute of Advanced Industrial Science and Technology (AIST),\\
	AIST Tsukuba Central 4, Tsukuba 305-8562, Japan}
\affiliation{Department of Applied Physics, University of Tokyo, Tokyo 113-8656, Japan}
\author{H. Takagi}
\affiliation{Department of Advanced Materials,
	University of Tokyo, Kashiwa 277-8561, Japan.}
\affiliation{Correlated Electron Research Center (CERC),\\
	National Institute of Advanced Industrial Science and Technology (AIST),\\
	AIST Tsukuba Central 4, Tsukuba 305-8562, Japan}
\affiliation{Advanced Science Institute, RIKEN (The Institute of Physical and
Chemical Research), Wako 351-0198, Japan}

\date{\today}
\begin{abstract}

Electrostatic carrier doping using a field-effect-transistor
structure is an intriguing approach to explore electronic phases by
critical control of carrier concentration \cite{RMP,Imada}.
We demonstrate the reversible control of  the insulator-metal transition (IMT)
in a two dimensional (2D) electron gas at the interface of insulating SrTiO$_3$
single crystals.
Superconductivity was observed in a limited number of devices doped
far beyond the IMT, which may imply the presence of 2D
metal-superconductor transition.
This realization of a two-dimensional metallic state
on the most widely-used perovskite oxide is the best manifestation of the potential
of oxide electronics.

\end{abstract}
\maketitle
The perovskite SrTiO$_3$ is sometimes called the ``silicon of oxide electronics''
because it is commonly used as a substrate for epitaxial growth of a variety
of oxide films \cite{Kawasaki}.
Not only it is an insulator with a band gap of 3.2\,eV but also a quantum paraelectric with
an extremely large dielectric constant of more than 10$^6$ at low temperatures \cite{Muller}.
N-type conduction has been realized by introducing a small number of oxygen vacancies
or by cation substitutions.
An insulator to metal transition occurs at a rather low carrier concentration
of $n \sim $10$^{18}$\,cm$^{-3}$ \cite{Tufte,Frederikse} for a doped oxide system,
which owes largely to the high mobility exceeding 10$^4$\,cm$^2$/Vs.
Superconductivity emerges in a limited concentration range from $n=7 \times 10^{18}$
to $5 \times 10^{20}$ cm$^{-3}$ and the transition temperature $T_c$ shows a bell 
shaped dependence on $n$ with an optimum
$T_c$\,$\sim$\,0.35\,K at $n$\,$\sim$\,$10^{20}$\,cm$^{-3}$ \cite{Koonce}.
The carrier concentration required to achieve superconductivity is
two orders of magnitude smaller than that for high-$T_c$ cuprates
which is a great advantage to realize electrostatic control
of insulator-metal and superconductor transitions.

Such a unique doping properties of SrTiO$_3$ has triggered attempts to 
dope charged carriers electrostatically by using a field effect 
transistor (FET) structure.
However, so far, attempts to realize metallic state have been
unsuccessful \cite{Ueno}.
This is due to several reasons, notably carrier trapping at the gate insulator/SrTiO$_3$
interface, Schottky barrier formation between SrTiO$_3$ and source and drain
electrodes, and insufficient breakdown voltage of the gate insulators.
An alternative approach has shown that two dimensional electron gas can be
created at SrTiO$_3$/LaAlO$_3$ hetero-interface by a charge-transfer due to a
polar discontinuity \cite{Ohtomo,Takahashi,Thiel,Herranz,Brinkman,Reyren}
or possible oxygen defects \cite{Maurice}.
This electronically tailored interface was found to be superconducting
as in the bulk material \cite{Reyren}.

Recently the performance of SrTiO$_3$ FET has been drastically
improved \cite{Shibuya2,Nakamura} by overcoming the contacts and interface problems.
These new devices showed metallic behavior down to 7\,K \cite{Nakamura}.
In this Letter, we extend our study to lower temperatures
below 1\,K, and show that the two-dimensional insulator-metal
transition is indeed achieved at the interface between parylene and SrTiO$_3$
reversibly while sweeping the gate voltage.
A superconducting transition to zero-resistance state was observed at 0.13\,K
only in a sample where a large enough number of carriers were successfully
introduced beyond the critical carrier number for the insulator-metal transition (IMT).
The observation of two dimensional insulator-metal-superconductor transition
represents major progress not only in the attempts for future
oxide electronics but also in oxide physics.

A top gate FET was fabricated consisting of Au/parelne/SrTiO$_3$ layers.
Al electrodes of the Hall bar geometry (Fig.1a) were evaporated on an atomically 
flat surface of SrTiO$_3$ (100) before depositing the gate insulator parylene.
The use of Al electrode is the key to suppressing Schottky barrier formation.
Details of the device fabrication are described elsewhere \cite{Nakamura}.
We have measured six samples (Sample A-F) with different parylene 
thickness (0.53-0.63\,$\mu$m), corresponding to capacitance $C_\mathrm{i}=4.4-5.3$\,nF/cm$^{2}$ 
allowing a wide-ranging threshold gate voltage ($V_\mathrm{th}$\,=\,97-170\,V).
Resistivity measurements above 2\,K were carried out by dc four-probe method with 
temperature control by the Quantum Design Physical Property Measurement System, 
while those below 1\,K were carried out by ac four-probe method (frequency 45\,Hz)
in a dilution refrigerator.

\begin{figure}[!htb]
	\centering
	 \includegraphics[width=\columnwidth,clip]{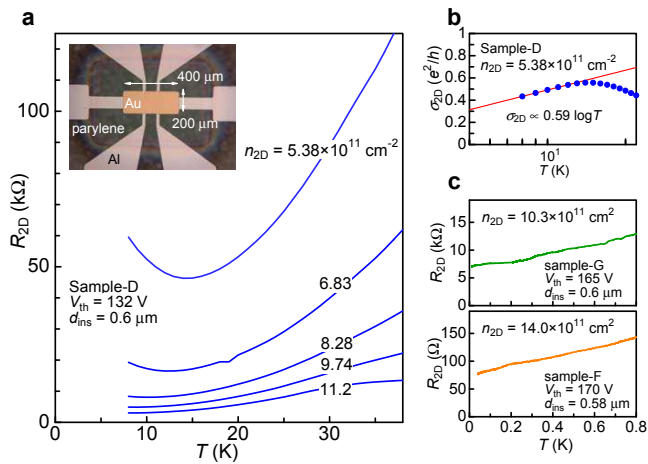} 
	\caption{%
Electrostatic control of metal-insulator transition of non-doped 
SrTiO$_3$ single crystal.
(a) Temperature dependence of the sheet resistance ($R_\mathrm{2D}$)
tuned by the gate voltage near the metal-insulator transition.
The parylene gate insulator thickness ($d_\mathrm{ins}$) was 0.6 $\mu$m.
Sheet carrier density ($n_{\mathrm{2D}}$) in units of $10^{11}$\,cm$^{-2}$
is shown.
$n_\mathrm{2D}$ is derived from the gate voltage $V_G$, capacitance 
$C_\mathrm{i}$ of parylene, and threshold gate voltage $V_\mathrm{th}$
($n_\mathrm{2D}=C_\mathrm{i}(V_\mathrm{G}-V_\mathrm{th})/e$).
The inset shows an photograph of sample with Al electrodes, parylene, and 
Au gate electrode.
Channel area is 400\,$\mu$m$\times$200\,$\mu$m.
(b) Sheet conductance in units of $e^2/h$ plotted against log$T$.
(c) Sheet resistance versus temperature at $T< 1$\,K in the case of 
$R_\mathrm{2D}<<h/e^2$.
Metallic behavior ($dR/dT > 0$) is seen down to the base temperature (15\,mK).
}
	\label{fig1}
\end{figure}

Figure 1a shows the temperature dependence of sheet
resistance for various gate voltages (or electron densities). 
As the gate voltage is increased, the temperature dependence changes from insulating
to metallic.
At high enough gate voltages, the resistance eventually becomes metallic all the
way down to a low temperature.
The sheet carrier density ($n_\mathrm{2D}$) here is estimated as
$C_\mathrm{i}(V-V_\mathrm{th})/e$, where $C_\mathrm{i}$ is the capacitance per unit area.  
The sheet carrier density ($n_\mathrm{2D}$) at the insulator to metal transition is
approximately \,1$\times$10$^{12}$\,cm$^{-2}$, which corresponds to a Fermi energy of about 1\,meV
assuming that the system is a two-dimensional (2D) free-electron gas.
We also note that the sheet resistance ($R_\mathrm{2D}$) in this regime is close to 
the two dimensional quantum sheet resistance $h/e^2$\,=25.8\,k$\Omega$.
In the insulating state, $R_\mathrm{2D}$ shows logarithmic temperature dependence
at lower temperatures as shown in Fig.\,1b, and the slope is of the order of
the quantum conductance.
At first glance these observations suggest two-dimensional 
weak-localization \cite{Bergmann,Kozuka}.
However, we show that the conduction in this weakly insulating regime
is highly likely to be percolative in nature and not described by the
weak localization scenario.

The above behavior suggests the emergence of 2D metal at the
parylene/SrTiO$_3$ interface.
This is in contrast to the scaling theory of localization, which predicts that
a 2D system should become insulating at zero temperature \cite{Abrahams}.
Recent experiments on very high mobility 2D electron gases formed in Si and GaAs hetero-structure,
however, demonstrated the existence of insulator-metal transitions even in 2D
and have been attracting a considerable attention \cite{Kravchenko}.
To explore the ground state of electrostatically induced metallic interface between
parylene and non-doped SrTiO$_3$ single crystal, we extend the transport
measurements down to 15\,mK using a dilution refrigerator for the samples with $R_{2D}$
well below the quantum resistance.
Figure\,1c shows a typical result.
The metallic behavior ($dR/dT>0$) is observed down to 
$\sim$15\,mK.
As shown in Fig.\,1a, when the gate voltage is reduced, $R_{2D}$ shows
an insulating ($dR/dT<0$) behavior.
We emphasize here that the two ground states can be tuned reversibly
by the electrostatic control through the parylene gate insulator.

\begin{figure}[!htb]
	\centering
 \includegraphics[width=\columnwidth,clip]{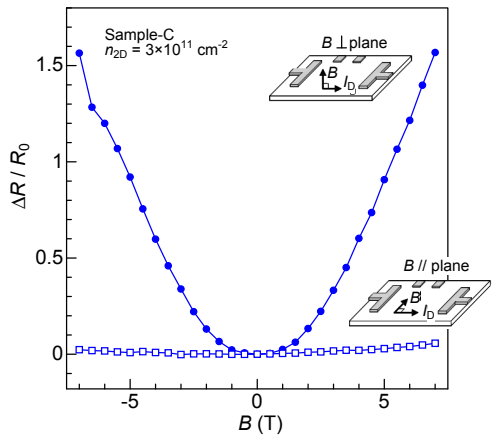}
	\caption{%
Anisotropic magnetoresistance.
Magnetoresistance $\Delta R/R_0$ for two different orientations of
magnetic field $B$ at 4.2\,K:
perpendicular and parallel fields to the SrTiO$_3$/parylene interface.
Positive magnetoresistance for the perpendicular field
is caused by the cyclotron motions of electrons and its magnitude
is related to the mobility $\mu$ of the electron gas ($\Delta R/R_0 \sim (\mu B)^2$).
Little effect for the parallel field indicates that the conducting layer
is much thinner compared to the cyclotron radius of electrons.
}
	\label{fig2}
\end{figure}

Compared to conventional semiconductors like Si or GaAs,
SrTiO$_3$ has a large dielectric constant of the order of $10^6$
at low temperatures \cite{Muller} which weakens the confining potential
at the interface.
Therefore, it should be clarified whether or not the electronic state formed at 
parylene/SrTiO$_3$ interface is two-dimensional.
We have checked this experimentally by measuring the anisotropy of
transverse magnetoresistance $\Delta R/R_0$.
Here, $\Delta R$ is the resistance change in a magnetic field 
and $R_0$ is the zero-field resistance.
Figure\,\ref{fig2} shows the magnetoresistance taken under $B$ perpendicular to the
plane, and parallel to the plane.
The large positive magnetoresistance, indicative of a high mobility,
is observed only for the field perpendicular to the plane.
No appreciable magnetoresistance is seen for the field parallel to the plane,
which implies that the thickness of the conducting layer is much thinner
than the cyclotron radius of electrons, $r_\mathrm{c}=\frac{m^{*}v_\mathrm{F}}{eB}=\frac{\hbar k_\mathrm{F}}{eB}$,
($m^{*}$; the electron effective mass \cite{Uwe}, $v_\mathrm{F}$; Fermi velocity, $k_\mathrm{F}$; Fermi wave number,
$\hbar = h/2 \pi$, where $h$ is Planck's constant), which we estimate $r_c$\,$\sim$15\,nm at $B=6$\,T.
Numerical calculation based on the Schr\"odinger-Poisson equation
with a triangular potential approximation estimates the
conducting layer depth to be approximately 10 nm in metal/insulator/SrTiO$_3$ 
structure \cite{Inoue} and support the presence of two-dimensional electron system.
It should also be noted that this estimation of conducting layer depth may become
even smaller when the rapid decrease of dielectric constant of SrTiO$_3$ at large
electric field is taken into account \cite{Christen,Susaki}.
From these results, we conclude that the metal-insulator transition observed
in the parylene/SrTiO$_3$-FET is two-dimensional in nature.

Figure 3a shows the sheet resistance $R_\mathrm{2D}$ as a function of the sheet carrier density
$n_\mathrm{2D}$ at 4.2\,K.
A notable feature of the electrostatically-induced IMT
is the sharp and almost discontinuous decrease of the resistance with increasing
the sheet carrier density.
With further increasing the carrier density, a ``kink'' in
$R_\mathrm{2D}$-$n_\mathrm{2D}$ is clearly recognized, and above the kink,
much more moderate decrease is observed,
where metallic behavior of resistivity is seen down to the lowest temperature.
The comparison of two kinds of mobilities deduced from the magnetoresistance 
and the field-effect, respectively, suggests phase separation into metallic and 
insulating domains in the transition region before reaching the kink,
implying that the insulator to metal transition is very likely discontinuous.
Figure 3b shows the evolution of carrier mobility near the transition, where
two independent mobilities are compared;
the field-effect mobility $\mu_{\mathrm{FE}} = (1/e)(d\sigma_\mathrm{2D}/dn_\mathrm{2D})$,
where $\sigma_\mathrm{2D}= 1/R_\mathrm{2D}$, $n_\mathrm{2D}=C_\mathrm{i}V_\mathrm{G}/e$,
and the magnetoresistance mobility $\mu_\mathrm{MR}$ estimated by the magnitude of the 
quadratic part of magnetoresistance $\mathit{\Delta} R/R_{0} =(\mu_\mathrm{MR}B)^2$ \cite{Jervis}.
While the two mobilities agree very well in the larger gate voltage region above the
kink, they are distinct from each other in the transition region (Fig. 3b).
The $\mu_\mathrm{MR}$ is as large as 2000 cm$^2$/Vs and surprisingly independent of
carrier density above and below the kink.
In contrast, $\mu_\mathrm{FE}$ shows a rapid decrease with decreasing voltage
around the transition region.
This contrasting behavior between the two mobilities may be understood naturally
in terms of phase separation.
The conduction should be dominated by the metallic domains.
The magnetoresistance is then almost identical to those of metallic domains
independent of the gate voltage.
On the other hand, the value of $\mu_\mathrm{FE}$ reflects the channel resistance.
Narrowing of the metallic domain during the separation enhances the
channel resistance and hence $\mu_\mathrm{FE}$ decreases rapidly.
Thus, the transition between metallic and insulating state tuned by a gate voltage
is very likely a first order transition accompanied by a phase separation.
The origin of seemingly weak localization behavior in the transition region
shown in Fig. 1 is not clear.
We suspect that narrow percolation path in the mixed phase might be
responsible for the apparent logarithmic temperature dependence.

\begin{figure}[!htb]
	\centering
    \includegraphics[width=\columnwidth,clip]{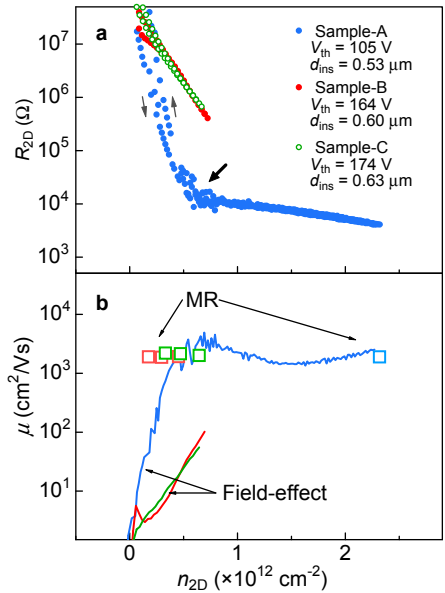}
	\caption{%
Electrostatic tuning of the sheet resistance and the mobilities.
(a) Sheet resistance $R_\mathrm{2D}$ plotted against the sheet carrier 
density $n_\mathrm{2D}$ at 4\,K.
The position of a ``kink'' is shown by a thick arrow.
(b) Field-effect mobility:
$\mu_{\mathrm{FE}} = (1/e)(d\sigma_\mathrm{2D}/dn_\mathrm{2D})$,
and magnetoresistance mobility $\mu_\mathrm{MR}$: $\mathit{\Delta} R/R_{0} =(\mu_\mathrm{MR}B)^2$.
Two mobilities show significant difference near the threshold
of gate voltage (early stage after the accumulation of mobile charge),
but coincide at large $n_\mathrm{2D}$.
}
	\label{fig3}
\end{figure}

The discontinuous transition presented here is reminiscent of the
``steplike'' change in conductance observed in SrTiO$_3$/LaAlO$_3$
hetero-structures as a function of LaAlO$_3$ thickness \cite{Thiel},
where the electric field produced by polar discontinuity is believed
to play a role.
In both cases, the sheet resistance of the conducting side of the
``discontinuous change'' is of order 10 kOhm, suggesting there exists a universal
minimum metallic conductivity of around $\sim 10^{-4}$\,S.
Although the mechanism of first-order-like metal-insulator transition is
currently unclear, the observation of similar behavior in different types
of device structure seems to indicate that the origin of discontinuous
metalization may be intrinsic at least to SrTiO$_3$.

In only a few samples, we were able to apply a large enough gate voltage without 
breaking the gate insulator (parylene) to go far beyond the discontinuous IMT.
With successful accumulation of an enough number of carriers, we
observed zero-resistance state associated with superconductivity.
As shown in Fig.\,4, at 128\,mK the resistance decreases gradually with
increasing gate voltage, namely the carrier concentration, and eventually
becomes zero within our resolution above $n_\mathrm{2D} \sim 2\times10^{12}$\,cm$^{-2}$.
With decreasing gate voltage, then, the resistance goes back to a finite
value reproducibly.
With increasing temperature from 128\,mK at $n_\mathrm{2D} \sim 2.2\times10^{12}$\,cm$^{-2}$,
the resistive transition to a normal state is observed around 130-160\,mK as seen
in the inset of Fig.\,4.
The application of magnetic field of 0.03\,T perpendicular to the plane was
found to suppress the zero resistance state.

\begin{figure}[!htb]
	\centering
	 \includegraphics[width=\columnwidth,clip]{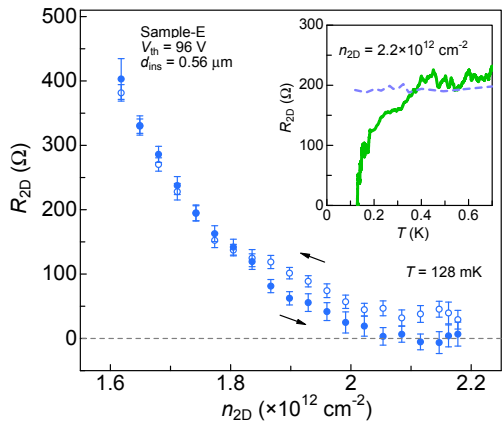}
	\caption{%
Continuous and reversible electrostatic tuning of superconductivity
at the surface of undoped SrTiO$_3$ single crystal.
Reversible and continuous control between the zero-resistance and the finite
resistance was achieved by sweeping the gate voltage at $T$\,=\,128\,mK.
The inset shows temperature dependence of the sheet resistance at a
constant sheet carrier density at $B$\,=0\,T (continuous line) and at
$B$\,=0.03\,T (dashed line).
Zero-resistance is seen below 130 mK for $B$\,=0\,T.
}
	\label{fig4}
\end{figure}

These observations demonstrate that superconductivity is marginally
achieved around $n_\mathrm{2D} \sim 2 \times10^{12}$\,cm$^{-2}$.
This sheet carrier density corresponds to $n_\mathrm{3D} \sim 2 \times10^{18}$\,cm$^{-3}$
assuming the thickness of 2D electron layer to be 10\,nm.
Though the reason is not obvious, this is similar to
the onset $n_\mathrm{3D} \sim 7 \times10^{18}$\,cm$^{-3}$ reported for bulk,
given the uncertainty in 2D electron layer thickness.
Note that we did not see any trace of superconductivity in the metallic sample shown
in Fig. 1c, down to 15\,mK at $n_\mathrm{2D} \sim 1\times10^{12}$\,cm$^{-2}$.
These suggest that the 2D insulator-metal transition occurs first and that, to achieve
2D metal-superconductor transition, extra charge carrier must be introduced.
We, however, cannot rule out completely superconductivity at an extremely
low temperature in the sample near IMT.

It should be noted that by applying a large gate field for a long period (typically
$V_\mathrm{G}$\,$>$\,150\,V for one week), another superconductor-like transition
without a zero resistance state appears around 0.35\,K which does not disappear
even after reducing the gate field.
This is highly likely to be the superconductivity associated with the creation
and/or migration of oxygen defects at the interface.

In conclusion, we have demonstrated that the parylene/SrTiO$_3$ interface shows a
transition from an insulator into two-dimensional metal by the electrostatic 
carrier doping as observed in high mobility 2D electron gas formed in Si and GaAs
hetero-structures.
We believe that this represents a breakthrough in oxide electronics.
One of the unique features of the parylene/SrTiO$_3$ interface is a
zero-resistance state emerging at 130\,mK.
The emergent superconductivity is thought-provoking not only because of the
possible 2D character but also because it develops in a electron gas with
a small Fermi energy of a few meV---usually trivial interactions like
spin-orbit coupling \cite{Winkler} can be comparable with Fermi energy
in this system and may give rise to an exotic superconductivity.
Therefore, what we have demonstrated in this study is not simply a hectic
chase of oxide electronics after the conventional semiconductor but
also a prologue to new paradigm of transition metal oxide physics.

We thank R. Perry for critical reading of the manuscript.
This work was supported in part by MEXT, Grant-in Aid for Scientific Research
(S)(19104008).


\begin{thebibliography}{99}

\bibitem{RMP}
C. H. Ahn \textit{et al}.,
Rev. Mod. Phys. {\bf 78}, 1185 (2006).

\bibitem{Imada}
M. Imada, A. Fujimori and Y. Tokura, 
Rev. Mod. Phys. {\bf 70}, 1039 (1998).

\bibitem{Kawasaki}
M. Kawasaki \textit{et al}.,
Science {\bf 266}, 1540 (1994).

\bibitem{Muller}
K. A. M\"uller and H. Burkard,
Phys. Rev. B {\bf 19}, 3593 (1979).

\bibitem{Tufte}
O. N. Tufte and P. W. Chapman,
Phys. Rev. {\bf 155}, 796 (1967).

\bibitem{Frederikse}
H. P. R. Frederikse and W. R. Hosler,
Phys. Rev. {\bf 161}, 822 (1967).

\bibitem{Koonce}
C. S. Koonce, Marvin L. Cohen, J. F. Schooley, W. R. Hosler
and E. R. Pfeiffer,
Phys. Rev. {\bf 163}, 380 (1967).

\bibitem{Ueno}
K. Ueno \textit{et al}.,
Appl. Phys. Lett. {\bf 83}, 1755 (2003).

\bibitem{Ohtomo}
A. Ohtomo and H. Y. Hwang,
Nature (London) {\bf 427}, 423 (2004).

\bibitem{Takahashi}
K. S. Takahashi \textit{et al}.,
Nature {\bf 441}, 195 (2006).

\bibitem{Thiel}
S. Thiel, G. Hammerl, A. Schmehl, C. W. Schneider, and J. Mannhart,
Science {\bf 313}, 1942 (2006).

\bibitem{Herranz}
G. Herranz, \textit{et al}.,
Phys. Rev. Lett. {\bf 98}, 216803 (2007).

\bibitem{Brinkman}
A. Brinkman, \textit{et al}.,
Nature Materials {\bf 6}, 493 (2007).

\bibitem{Reyren}
N. Reyren, \textit{et al}..
Science {\bf 317}, 1196 (2007).

\bibitem{Maurice}
J. L. Maurice \textit{et al}.,
Europhysics Letters {\bf 82}, 17003 (2008).

\bibitem{Shibuya2}
K. Shibuya, \textit{et al}.,
Appl. Phys. Lett. {\bf 88}, 212116 (2006).

\bibitem{Nakamura}
H. Nakamura \textit{et al}.,
Appl. Phys. Lett. {\bf 89}, 133504 (2006).

\bibitem{Bergmann}
G. Bergmann,
Phys. Rep. {\bf 107}, 1 (1984). 

\bibitem{Kozuka}
Y. Kozuka, Y. Hikita, T. Susaki and H. Y. Hwang,
Phys. Rev. B {\bf 76}, 085129 (2007).

\bibitem{Abrahams}
E. Abrahams, P. W. Anderson, D. C. Licciardello, and T. V. Ramakrishnan,
Phys. Rev. Lett. {\bf 42}, 673 (1979).

\bibitem{Kravchenko}
S. V. Kravchenko, and M. P. Sarachik,
Rep. Prog. Phys. {\bf 67}, 1 (2004).

\bibitem{Inoue}
I. H. Inoue,
Semicond. Sci. Technol. {\bf 20}, S112 (2005).

\bibitem{Christen}
H.-M. Christen, J. Mannhart, E. J. Williams, and Ch. Gerber,
Phys. Rev. B {\bf 49}, 12095 (1994).

\bibitem{Susaki}
T. Susaki, Y. Kozuka, Y. Tateyama, and H. Y. Hwang,
Phys. Rev. B {\bf 76}, 155110 (2007).

\bibitem{Uwe}
H. Uwe, R. Yoshizaki, T. Sakudo, A. Izumi,
and T. Uzumaki,
Jpn. J. Appl. Phys. {\bf Suppl.24-2}, 335 (1985).

\bibitem{Jervis}
T. R. Jervis and E. F. Johnson,
Solid-State Electron. {\bf 13}, 181 (1970).

\bibitem{Winkler}
R. Winkler, \textit{Spin-orbit coupling effects in two-dimensional
electron and hole systems.}, (Springer, Berlin, 2003).

\end{thebibliography}
\end{document}